\documentclass[aps,prl,twocolumn,showpacs,superscriptaddress,groupedaddress]{revtex4}  
\usepackage{graphicx}  
\usepackage{dcolumn}   
\usepackage{bm}        
\usepackage{amssymb}   
\usepackage{color}
\usepackage{amsmath}
\newcommand{\pd}{\partial}

\begin{document}

\widetext

\begin{flushright}
\end{flushright}

\title{
Path Integral Optimization for $T\overline{T}$ Deformation}
\author{Ghadir Jafari}
\email{ghjafari@ipm.ir}

\affiliation{\it School of Particles and Accelerators, Institute for Research in Fundamental Sciences (IPM), P.O.Box 19395-5531, Tehran, Iran}

\author{Ali Naseh} \email{naseh@ipm.ir}

\affiliation{\it School of Particles and Accelerators, Institute for Research in Fundamental Sciences (IPM), P.O.Box 19395-5531, Tehran, Iran}

\author{Hamed Zolfi} \email{hamed.Zolfi@physics.sharif.edu}

\affiliation{\it Department of Physics, Sharif University of Technology, P.O. Box 11365-9161, Tehran, Iran}\affiliation{\it School of Particles and Accelerators, Institute for Research in Fundamental Sciences (IPM), P.O.Box 19395-5531, Tehran, Iran}

\date{\today}

\begin{abstract}
We use the path integral optimization approach of Caputa, kundu, Miyaji, Takayanagi and Watanabe to find the time slice of geometries dual to
vacuum, primary and thermal states in the $T\overline{T}$ deformed two dimensional CFTs. The obtained optimized  geometries actually capture the entire bulk which fits well with the integrability and expected
UV-completeness of $T\overline{T}$-deformed CFTs. When deformation parameter
is positive, these optimized solutions can be reinterpreted as geometries at
finite bulk radius, in agreement with a previous proposal by McGough, Mezei and Verlinde. We also calculate the holographic entanglement entropy and quantum state complexity for these solutions. We show that the complexity of formation for the thermofield double state in the deformed theory is UV finite and it depends to the temperature. \end{abstract}

\maketitle
\section{Introduction}
\vspace{-.45cm}
Applications  of  quantum  information concepts to gravity and high energy physics recently
led to many wide-spreading developments.  In particular quantum  entanglement  helps  us to  understand how  gravity  emerges from field  theories \cite{Ryu2006,VanRaamsdonk}. One concrete idea to realize this emergent spacetime is to utilize the connection between tensor networks and holographic entanglement surface. It is argued that a time slice of AdS spacetime correspond to a special tensor network called multi-scale entanglement renormalization ansatz (MERA) \cite{Swingle}. However, when one tries to understand the interior of black hole, the entanglement is not enough \cite{Susskind2014}. This observation has led to significant interest to the information notion of quantum state complexity which can be used as a probe to investigate the growth rate of the Einstein-Rosen bridge \cite{EREPR}. 

Two holographic conjectures are proposed for quantum complexity: the CV conjecture \cite{susskind2014, Alishahiha} and CA conjecture \cite{susskind2015} which share some similarities but differ in many ways. To understand better the holographic results, recently the computational complexity for quantum field theory states have been studied \cite{myers2017,heller2017}. In one approach which is based on the idea of Nielsen and collaborators \cite{nielsen}, one associates a geometry to the space of unitaries that connect the desired state to a reference state. Then the complexity of desired state is defined as a length of a geodesic in this geometry. Up to now, this approach is well established just for Gaussian states. In another approach which is applicable for any 2D CFTs (free or interacting) \cite{PI1,PI2,PI3,PI4}, one works with Euclidean path integral description of quantum state and perform the optimization by changing the structure (or geometry) of lattice regularization. The resulting change in path integral-namely the Liouville action, was defined as a complexity of the corresponding state. Very interestingly, it has been shown recently that these two seemingly different approaches are surprisingly connected \cite{Caputa, heller2019}.

In this letter we would like to apply the path integral optimization \cite{PI2}, to the new class of integrable quantum field theories.
These theories are introduced by Smirnov and Zamolodchikov \cite{Zamolodchikov1,Zamolodchikov2} which  they studied the deformation of 2D integrable QFTs by a special irrelevant composite operator $T\overline{T}$, and found that the theory remains integrable even after the deformation. Interestingly, the resulting theory is UV complete and non-local and its energy spectrum and the S-matrix can be found exactly. After the proposal of \cite{verlinde}, which relates the $T\overline{T}$ deformation of 2D CFTs to 3D AdS gravity with a finite bulk cut-off, these new integrable QFTs have received remarkable attention in theoretical high energy  community. 

In the following, we show that the outcome of path integral optimization for $T\overline{T}$ deformation of 2D CFTs is geometries which they captures the entire bulk, specially the region out side the finite bulk cut-off in \cite{verlinde}. This result is in agreement with recent study \cite{Guica} where it is shown that the dual gravitational theory to the $T\overline{T}$ deformed 2D CFTs, has mixed boundary conditions for the non-dynamical graviton. It is worth to mention that our result for positive sign of deformation parameter can be interpreted as existence of hard bulk cut-off in agreement with \cite{verlinde}. Moreover, we calculate the holographic entanglement entropy for the obtained solutions and also quantum state complexity for some dual boundary QFT states. These states are $T\overline{T}$ deformation of vacuum, primary and thermal states. The quantum complexity of these states has the same UV structure as corresponding states in 2D CFTs. Moreover, the complexity of formation for thermofield double state is finite and it depends to the temperature \footnote{Based on bulk cut-off interpretation, the holographic complexity (CA) of thermofield double state in $T\overline{T}$ deformed CFTs is studied in \cite{Akhavan}.}.  
\section{Path Integral Optimization}
For 2D CFTs, the ground state wave functional on $\mathbb{R}^{2}$ is computed by an Euclidean path integral:
\begin{align}
\label{PIn}
\Psi_{\text{CFT}}[\bar{\Phi}(x)]=&\int\left(\prod_{x}\prod_{-\infty\leq \tau<\epsilon} \mathcal{D}\Phi\right) e^{-S_{\text{CFT}}(\Phi)}\times
\nonumber\\
&\times\prod_{x}\delta\left({\Phi}(\epsilon,x)-\bar{\Phi}(x)\right),
\end{align}
where $\tau$ is Euclidean time and $\epsilon$ is UV cutoff (i.e. the
lattice constant). It is worth noting that to evaluate this
discretized path integral in an optimal way, one can omit
any unnecessary lattice sites. To systematically quantify
such coarse-graining, one might introduce a 2D metric
(on which the path integration is performed)
\begin{equation}
\label{Fmet}
ds^{2} = \frac{l^2}{\epsilon^{2}}\left(d\tau^{2}+dx^{2}\right),
\end{equation}
such that one lattice site has a unit area. The optimiza-
tion procedure then can be described by modifying the
background metric for the path integration as
\begin{equation}
ds^{2} = g_{\tau\tau}(\tau,x)d\tau^{2}+g_{xx}(\tau,x)dx^{2}+2g_{\tau,x}(\tau,x)d\tau dx.
\end{equation}
Now, the key point is that the optimized wave functional,
up to a normalization factor, should be proportional to
the correct ground state wavefunction, i.e. the (\ref{PIn}) for the
metric (\ref{Fmet}). This constraint implies that
\begin{equation}
g_{\tau\tau}(\epsilon,x) = g_{xx}(\epsilon,x)=\frac{l^2}{\epsilon^{2}},\hspace{.5cm}g_{\tau,x}(\epsilon,x)=0.
\end{equation}
For 2D $T\overline{T}$ deformed CFTs, which are described by
the action
\begin{equation}
\label{STTb}
S = S_{\text{CFT}} +\mu \int d\tau dx \hspace{.5mm} T\overline{T}(\tau,x),
\end{equation}
the deformation operator is a double-trace operator. According to \cite{Witten}, this double-trace deformation
of CFT corresponds to imposing mixed boundary condition on the bulk metric. Motivating by this fact and noting that the 
position dependent coupling $\mu$ is related to the bulk matter field, in the following we keep the
coupling constant $\mu$ independent from ($\tau,x$) coordinates
and just allow the metric to vary. Accordingly, the optimization
is performed by the following ansatz
\begin{equation}
\label{Cmet}
ds^{2} = e^{2\Omega(\tau,x)} \left(d\tau^{2}+dx^{2}\right),
\end{equation}
where that mixed boundary condition is encoded in two requirements:  $e^{2\Omega(\epsilon,x)} =\frac{l^2}{\epsilon^{2}}$ and the energy of optimized solution matches with the energy spectrum of $T\overline{T}$ deformed CFT. 

On the metric $\hat{g}_{\alpha\beta} = e^{2\Omega}(\tau,x) g_{\alpha\beta}$, the partition function of 2D $T\overline{T}$ deformed CFTs is given by
\begin{align}
Z_{\mu}[\hat{g}] = \int \mathcal{D} &\Phi\hspace{1mm} \text{Exp} \bigg[-S_{\text{CFT}}[\Phi,\hat{g}] \hspace{.5mm}-
\nonumber\\
&-\int d^{2}\sigma \sqrt{\hat{g}} \left(\Lambda +\mu T\overline{T}\hspace{.1mm}[\Phi,\hat{g}]\hspace{.5mm}\right)\bigg],
\end{align}
where $d^{2} \sigma = d\tau dx$ and $\Lambda$ is related to the the cosmological counterterm. It is worth noting that this counterterm action is responsible for removing the UV divergence in presence of $T\overline{T}$ deformation. Under change of Weyl factor $\Omega(\tau, x)$,
this partition function changes to
\begin{align}
\label{DefPF}
&\frac{1}{Z_{\mu}[\hat{g}]} \frac{\partial Z_{\mu}[\hat{g}]}{\partial\Omega} = \frac{1}{Z_{\mu}[\hat{g}]}\int \mathcal{D}\Phi\hspace{.5mm} e^{-S_{\text{CFT}}-\int d^{2}\sigma \sqrt{\hat{g}} \left(\Lambda +\mu T\overline{T}\right)}\times\nonumber\\
&\hspace{.7cm}\times \left(-\frac{\pd S}{\pd \hat{g}_{\alpha\beta}}\frac{\pd \hat{g}_{\alpha\beta}}{\pd \Omega}- \int d^{2}\sigma \frac{\pd}{\pd \Omega} \sqrt{\hat{g}}\hspace{.5mm}(\Lambda+\mu T\overline{T})\hspace{.5mm}\right)
\end{align}
Using the above equation and also assuming that $\mu$ is
small, we get
\begin{equation}
\frac{\pd \log Z_{\mu}[\hat{g}]}{\pd \Omega} = \frac{c}{24\pi} \sqrt{\hat{g}} \hat{R}[\hat{g}]-2\Lambda \sqrt{\hat{g}}-\mu\hspace{.5mm}\pd_{\Omega} \langle T\overline{T}\rangle_{\hat{g}},
\end{equation}
where constant $c$ is the central charge of 2D CFT and 
\begin{equation}
\label{ExpTTb}
\langle T\overline{T}\rangle_{\hat{g}}=\hspace{-1mm} \int\hspace{-1mm} d^{2}\hspace{-1mm}\sigma \sqrt{\hat{g}}\hspace{.5mm}\hspace{.2mm} \big(\langle0|T_{zz}|0\rangle_{\hat{g}}\langle 0|\overline{T}_{\bar{z}\bar{z}}|0\rangle_{\hat{g}}-e^{-4\Omega}\langle 0|T_{z\bar{z}}|0\rangle_{\hat{g}}^{2}\hspace{.5mm}\big).
\end{equation}
One can now just treat (\ref{DefPF}) as a differential equation for the partition function $Z_{\mu}$ and solve it. This allows us to express the partition function $Z_{\mu}[\hat{g}]$, defined on one metric $\hat{g}$, in terms of
$Z[g]$, defined on metric $g$. The relationship is,
\begin{equation}
\label{Zmu}
Z_{\mu}[\hat{g}] = e^{S_{GL}[\Omega,g]-S_{GL}[0,g]}\hspace{.5mm}Z_{\mu}[g],
\end{equation}
where
\begin{align}
\label{GLA}
S_{GL}[\Omega,g] =\hspace{-.1cm}\int d^{2}\sigma\hspace{.5mm} \bigg(\frac{c}{24\pi}\sqrt{g}&\left(R[g]\Omega -\Omega \nabla^{2}\Omega -\bar{\Lambda}e^{2\Omega}\hspace{.5mm}\right)-
\nonumber\\
&-\mu\langle T\overline{T}\rangle_{\hat{g}}\bigg),
\end{align}
and $\bar{\Lambda} = \frac{24\pi}{c}\Lambda$. The subscript (GL) in the above equation means generalize Liouville since the first term in the above equation is actually the standard Liouville action. For $g_{\alpha\beta} =\delta_{\alpha\beta}$, the last term in (\ref{GLA}) can be calculated explicitly by noting that in complex coordinates ($w=\tau+ix$ , $\bar{w}=\tau-ix$) we have the below relations 
\begin{align}
\label{ExpV}
&\langle 0|T_{zz}|0\rangle_{e^{2\Omega}\delta_{\alpha\beta}}=\dfrac{c}{12}\hspace{.3mm}e^{-2\Omega} {\partial^{2}\Omega},
\nonumber\\
&\langle 0|\overline{T}_{\bar{z}\bar{z}}|0\rangle_{e^{2\Omega}\delta_{\alpha\beta}}=\dfrac{c}{12}\hspace{.3mm} e^{-2\Omega}\bar{\partial}^{2}\Omega,
\nonumber\\
&\langle 0|T_{z\bar{z}}|0\rangle_{e^{2\Omega}\delta_{\alpha\beta}}=\dfrac{c}{6}\hspace{.3mm} {\partial\bar{\partial}\Omega}.
\end{align}
By substituting (\ref{ExpV}) in (\ref{ExpTTb}), the action (\ref{GLA}) simplified to 
\begin{align}
\label{SGL}
&S_{GL}[\Omega,\delta]=\dfrac{c}{24\pi}\int d^{2}\omega\bigg[-4\Omega\partial\bar{\partial}\Omega-\tilde{\Lambda} e^{2\Omega}
\nonumber\\
&\hspace{.7cm}-\tilde{\mu} e^{-2\Omega}\left(\partial^{2}\Omega\bar{\partial}^{2}\Omega-4(\partial\bar{\partial}\Omega)^{2}+\dfrac{3}{16}\tilde{\Lambda}^{2}e^{4\Omega}\right)\bigg], 
\end{align}
where $\tilde{\mu} = \mu\pi c/6$. In obtaining the above equation, we used $\bar{\Lambda} = \tilde{\Lambda}+\frac{3}{16}\tilde{\mu}\tilde{\Lambda}^{2}$ where the reason for that will be clear in the following. Finally, the definition (\ref{PIn}) together with (\ref{Zmu}) imply that the ground-state wave functional $\Psi_{e^{2\Omega}\delta_{\alpha\beta}}$
computed from the path integral for the metric (\ref{Cmet}) is proportional to the one $\Psi_{\delta_{\alpha\beta}}$
for the flat metric 
\begin{equation}
\Psi_{e^{2\Omega}\delta_{\alpha\beta}}(\bar{\Phi}(x)) = e^{S_{GL}[\Omega,\delta]-S_{GL}[0,\delta]} \Psi_{\delta_{\alpha\beta}}(\bar{\Phi}(x)).
\end{equation}
It is worth to emphasize that these two states describe the same quantum state if at UV cutoff, $\Omega(\epsilon,x) = \log(l/\epsilon)$. 
\section{Optimizing Various States in $T\overline{T}$ deformed CFTs}
Now, according to \cite{PI1,PI2} the optimization is equivalent to
minimizing the normalization factor $e^{S_{GL}[\Omega,\delta]}$
of the wave functional. The intuition behind this proposal comes from the tensor network representation of vacuum wave functional. In that language, a quantum state is defined as a minimal number of the quantum gates (operators) needed to create the state starting from a reference state. 
Here, the factor $e^{S_{GL}[\Omega,g]}$ actually measures the number of repetitions of the same operation (i.e. the path
integral over a cell). In order to find this minimum value, we must vary the action $S_{GL}[\Omega,\delta]$ (\ref{SGL})
, which gives following equation  of motion:
\begin{align}
\label{eom1}
&4 \pd\bar{\pd}\Omega-\tilde{\Lambda} e^{2 \Omega}+\tilde{\mu}\hspace{.5mm}\bigg( \dfrac{3}{16}\tilde{\Lambda}^{2} e^{2 \Omega }+3e^{-2 \Omega } \pd^2\bar{\pd}^2\Omega\hspace{.5mm}+
\nonumber\\
&\hspace{-.15cm}+e^{-2\Omega}\bar{\pd}^2\Omega \hspace{.5mm}\big(\hspace{-.5mm}-2(\pd\Omega)^2+3\pd^2\Omega\big)
-2e^{-2 \Omega } \bigg[3\pd\Omega\hspace{.5mm} \pd\bar{\pd}^2\Omega\hspace{.5mm}+\nonumber\\
&\hspace{-.15cm}+(\bar{\pd}\Omega)^{2}\partial^{2}\Omega
+6 (\pd\bar{\pd}\Omega)^2\hspace{-.5mm}+\bar{\pd}\Omega\big(\hspace{-.5mm}-8\pd\Omega\hspace{.5mm}\pd\bar{\pd}\Omega+3\pd^2\bar{\pd}\Omega\big)\bigg]\bigg)\hspace{-.1cm}=0
\nonumber\\
\end{align}
Since finding the analytic solutions of this equation in general is difficult, one can solve it perturbatively in $\tilde{\mu}$ parameter. As in this letter, we are interested to find the solutions which are only depend to $2\tau= (\omega+\bar{\omega})$, we consider $\Omega(\omega+\bar{\omega}) = \Omega_{\text{CFT}}(\omega+\bar{\omega}) + \tilde{\mu}\hspace{.5mm} \Omega_{T\overline{T}}(\omega+\bar{\omega})$. According to \cite{PI4}, a vacuum state in two dimensional QFT can be constructed from various 
co-dimension one surfaces in the 3D gravity dual. For example one of them can be 2 dimensional  boundary and another can be time slice of 3D bulk geometry. These two surfaces can be related with the bulk coordinate transformations and till they have the same topology, they give the same state up to the normalization factor. In the following, we represent this bulk co-dimension one surface with $(z,x)$ coordinates.

Now we are going to solve (\ref{eom1}). If we set $\tilde{\mu}=0$, the solution which minimize $S_{GL}(\Omega,\delta)$ (\ref{SGL}) is
\begin{equation}
\label{sol1}
\Omega_{\text{CFT}} = -\frac{1}{2} \log \left(\tilde{\Lambda} z^2\right)
\end{equation}
which  describes the time slice of AdS$_{3}$ geometry in Poincare coordinate. Using this solution for unperturbed CFT$_{2}$, the first order perturbation $\Omega_{T\overline{T}}$ becomes
\begin{equation}
\label{Psol1}
\Omega_{T\overline{T}}(z) = c_1 \frac{z^2}{l^2}+c_2\frac{l}{z},
\end{equation}
where $c_1$ and $c_2$ are arbitrary dimensionless integration constants. By choosing  $\tilde{\Lambda} =1/l^{2}$, the optimized metric (\ref{Cmet}) becomes
\begin{equation}
\label{AdsDef1}
ds^2=\frac{l^2}{z^2}\left(1+8c_{1}\tilde{\mu}\hspace{1mm}\frac{z^{2}}{l^2} +c_{2}\tilde{\mu}\hspace{1mm}\frac{l }{z}\right)(dz^2+dx^2).
\end{equation}
Imposing the UV condition $g_{zz}(z=\epsilon,x) \sim 1/\epsilon^{2}$ and IR condition $g_{zz}(z=\infty,x)=0$ respectively imply that $c_{2}$ and $c_1$ should be zero which it means that the Poincare AdS$_3$ remains undeformed.
Another interesting solutions for $\tilde{\mu}=0$ are excited states created
by acting a primary operator
$\mathcal{O}_{\alpha}$ with the conformal dimension $h_{\alpha}=\bar{h}_{\alpha}$ which its behavior under the Weyl re-scaling is expressed as
\begin{equation}
\mathcal{O}_{\alpha} \sim e^{-2h_{\alpha}}\mathcal{O}_{\alpha}.
\end{equation}
It is shown that in absence of $\tilde{\mu}$, the geometry dual to this state is given by
\begin{equation}
\Omega_{\text{CFT}}=\frac{1}{2}\log\left(\dfrac{a^2}{\sinh^{2}(\frac{az}{l})}\right),\hspace{.5cm} a=1-\dfrac{12h_{\alpha}}{c}
\end{equation}
For $a=1$, this describes the time slice of global AdS$_{3}$. Using this unperturbed solution in (\ref{eom1}), the perturbed solution becomes
\begin{align}
\label{PS}
ds^2\hspace{-.2mm}=\hspace{-.2mm}a^2\text{csch}^2(\frac{a z}{l})\bigg( \hspace{-.5mm}1+\frac{2\tilde{\mu}c_1}{l^2}\big(1-\frac{a z}{l} \coth (\frac{a z}{l})\big)\hspace{-.7mm}\bigg)(dz^2\hspace{-.2mm}+\hspace{-.2mm}dx^2),
\end{align}
where $ c_1$ is a dimensionless constant of integration and also a coefficient of a term which violates the condition $g_{zz}(\epsilon,x)\sim\frac{1}{\epsilon^{2}}$ is set to zero. 
Using the coordinate transformation 
\begin{equation}
\label{FGPS}
z=\frac{2l}{a} \tanh ^{-1}(\frac{\sqrt{\rho }}{2})\big(1- \frac{\tilde{\mu} c_1}{l^2}\big),
\end{equation}
the metric (\ref{PS}) takes below Fefferman-Graham expansion
\begin{align}
\label{FGPS2}
ds^2=\frac{l^2 a^2}{4\rho^2}d\rho^2 +\frac{l^2 a^2}{\rho}(1+\frac{2 \tilde{\mu}c_1}{l^2})\big(1 -\frac{1}{2}\rho+
\frac{1}{16}\rho^2)dx^{2}.
\end{align}
The gravitational energy of the this solution matches with correspondent energy of deformed primary state \cite{Zamolodchikov1} for $ c_1=-1/32\pi^2$.

All we have done up to now was for a $T\overline{T}$ deformation of a CFT at zero temperature. To extend the analysis to a finite temperature $T =1/ \beta^{\prime}$
case, one can use the thermofield double representation of wave functional. In this representation, the wave functional is computed from a path integral on a cylinder with a finite width $-\frac{\beta^{\prime}}{4}< z<\frac{\beta^{\prime}}{4}$ accordingly
\begin{align}
\label{PInTer}
&\Psi_{\beta^{\prime}}[\bar{\Phi}_2(x),\bar{\Phi}_1(x)]=\int\left(\prod_{x}\prod_{-\frac{\beta^{\prime}}{4}\leq z<\frac{\beta^{\prime}}{4}} \mathcal{D}\Phi\right) e^{-S_{}(\Phi)}\times
\nonumber\\
&\times\prod_{x}\delta\left({\Phi}(-\frac{\beta^{\prime}}{4},x)-\bar{\Phi}_1(x)\right)\delta\left({\Phi}(\frac{\beta^{\prime}}{4},x)-\bar{\Phi}_2(x)\right),
\end{align}
where $S$ is given by (\ref{STTb}). In absence of deformation, it is shown \cite{PI1}
that the minimization procedure gives
\begin{equation}
\label{OmegaCFT}
\Omega_{\text{CFT}}(z)=\frac{1}{2}\log \left(\frac{4\pi^2 l^2 \sec^2\left(\frac{2\pi  z}{\beta
}\right)}{\beta ^2}\right),
\end{equation}
which is actually describe time slice of BTZ black hole. It is worth noting that the temperature for the solution (\ref{OmegaCFT}) is assumed to be different with the temperature $1/\beta^{\prime}$. The reason for this will be clear in the following. Substituting this solution in (\ref{eom1}) gives following equation for $\Omega_{T\overline{T}}$,
\begin{equation}
\label{s1}
\frac{\beta^2}{4} \cos ^2\left(\frac{2\pi   z}{\beta}\right)\pd_{z}^{2}\Omega _{T\overline{T}}(z)-2 \pi^2  \Omega_{T\overline{T}}(z)=0,
\end{equation}
which its solution is
\begin{equation}
\label{s2}
\Omega_{T\overline{T}}(z) =\left(\frac{2\pi c_1  z}{\beta }
\right) \tan \left(\frac{2\pi  z}{\beta
}\right)+c_1.
\end{equation}
In the above expression $c_1$  is arbitrary integration constant. Substituting (\ref{s1}) and (\ref{s2}) in (\ref{eom1}), up to first order in $\tilde{\mu}$, we have
\begin{align}
\label{BTZ1}
ds^{2}=&\dfrac{4\pi ^2 l^2 }{\beta^2}\sec ^2(\frac{2\pi z}{\beta
})\bigg(1+\frac{2\tilde{\mu}c_1}{l^2}\big[1+
\frac{2\pi
z}{\beta} \tan(\frac{2\pi z}{\beta})\big]\bigg)\times
\nonumber\\
&\hspace{.8cm}\times(dz^2+dx^2).
\end{align}
To determine the constant $c_1$, it is better to transform this metric in the global coordinate. This can be achievable according the below coordinate transformation
\begin{align}
z=\frac{\beta}{2\pi }(1-\frac{\tilde{\mu} c_1}{l^2})\cos ^{-1}(\frac{2\pi l^2}{\beta  r})-\frac{\tilde{\mu} c_1 }{r \sqrt{1-\frac{4\pi^2 l^4}{\beta ^2 r^2}}},
\end{align}
which by that the metric (\ref{BTZ1}) changes to
\begin{align}
\label{BTZ2}
\hspace{-.1cm}ds^{2} = \frac{dr^{2}}{f(r)} +r^{2} dx^{2},\hspace{.1cm}
f(r)=\frac{r^2}{l^2}-\frac{4 \pi^2 l^{2}}{\beta ^2}(1+\frac{2\tilde{\mu}}{l^2}c_1).
\end{align}
The temperature of this black hole is
\begin{align}
\label{TTTb}
1/\beta^{\prime}\equiv T_{T\overline{T}} = (1+\frac{\tilde{\mu}}{l^2}c_1)/\beta.
\end{align} Let us remind that the time slice of standard BTZ black hole with the ADM mass $M$ is
\begin{equation}
\label{BTZ}
ds^{2} = \frac{dr^{2}}{f(r)} +r^{2} dx^{2},\hspace{.16cm}
f(r)=\frac{r^2}{l^2}-8G M_{\text{BTZ}},
\end{equation} 
where the mass and temperature of it is related according to following
\begin{equation}
\label{MBTZ}
 M_{\text{BTZ}} = \frac{\pi^2 l^2}{2G}\frac{1}{\beta^{2}}.
\end{equation}
For $c_{1} = G M_{\text{BTZ}}/4\pi^{2}$ and considering the $T_{T\overline{T}}$ in the right hand side of (\ref{MBTZ}) instead of $1/\beta$, for $M_{T\overline{T}}$ up to first order in $\mu$, we find 
\begin{equation}
\label{MTTb}
M_{T\overline{T}} = M_{\text{BTZ}} \hspace{.5mm}(1+ \frac{M_{\text{BTZ}}}{8\pi l}\mu),
\end{equation}
which exactly matches with the energy spectrum of $T\overline{T}$ deformed CFT \cite{Zamolodchikov1,Zamolodchikov2} for small $\mu$. By using
\begin{align}
r(\rho) = -\frac{l(\beta^{2}+\pi^2l^2\rho )}{\beta^2 \sqrt{\rho}} \left(1+c_1 \frac{\tilde{\mu}}{l^2}\right),
\end{align}
the metric (\ref{BTZ2}) changes to
\begin{align}
\label{BTZ3}
ds^2&=\dfrac{l^2 d\rho^2}{4\rho^2}+\frac{l^2(1+\frac{2\tilde{\mu}}{l^2}c_1)}{\rho}\bigg(1+\frac{2\pi^2l^2}{\beta^2}\rho+\frac{\pi^{4}l^4}{\beta^4}\rho^{2}\bigg)dx^{2}.
\end{align}
To understand better this result, let us remind that The most general solution of 3 dimensional Einstein’s equations with a negative cosmological constant can be expressed in Fefferman–Graham gauge by
\begin{align}
\label{FG}
& ds^{2} =\frac{l^{2}}{4\rho^{2}}d\rho^{2}+\frac{l^{2}}{\rho}g_{\alpha\beta}(\rho,x) dx^{\alpha}dx^{\beta},
\nonumber\\
&g_{\alpha\beta}(\rho,x^{\alpha}) = g^{(0)}(x^{\alpha}) +g^{(2)}_{\alpha\beta}(x^{\alpha})\rho + g^{(4)}_{\alpha\beta}(x^{\alpha})\rho^{2},
\end{align}
which $g^{(4)}$ and $g^{(2)}$ are determined algebraically in terms of $g^{(0)}$ \cite{skenderis}. Based on variational principle approach to the $T\overline{T}$
deformation of CFTs in \cite{Guica} (actually with mixed boundary condition as we discussed below Eq.(\ref{STTb})), it is argued that in the deformed theory the source for the stress tensor is given by a non-linear combination of the metric and stress tensor expectation value in the original CFT as following \footnote{We have considered our different convention in defining the $T\overline{T}$ operator with \cite{Guica} in coefficient of $g^{(2)}$ term.}
\begin{align}
g^{[\mu]}_{\alpha\beta} = g^{(0)}_{\alpha\beta} +\frac{\mu }{32\pi Gl}\hspace{.5mm} g^{(2)}_{\alpha\beta} +\mathcal{O}(\mu^{2}).
\end{align}
For the BTZ black hole $g^{(0)}_{xx}=1$ and $g^{(2)}_{xx} = \frac{2\pi^{2}l^2}{\beta^{2}}$, which together with (\ref{MBTZ}) implies that
\begin{equation}
g^{[\mu]}_{xx} = 1+\frac{M_{\text{BTZ}}}{8\pi l}\mu+\mathcal{O}(\mu ^{2}).
\end{equation}
Remarkably, it is the conformal boundary of (\ref{BTZ3}) by setting $c_{1} = G M_{\text{BTZ}}/4\pi^{2}$ same as previous analysis for the energy of black hole solution (\ref{BTZ2}). Another interpretation for the geometry (\ref{BTZ3}) is that its conformal boundary corresponds to fixing the induced metric on a constant
$\rho =\rho _{c}$ surface, with
\begin{align}
\label{rcut}
\rho_{c} = \frac{\mu }{32\pi G l},	
\end{align}
in agreement with the earlier proposal of \cite{verlinde}. It is worth noting that the path integral optimization solution (\ref{BTZ2}) shows that the entire spacetime should be kept especially
the region outside the would-be “cutoff” surface, which in proposal \cite{verlinde} is removed. Since $T\overline{T}$-deformed CFTs are conjectured to contain (non-local) observables of arbitrarily high energy, it makes sense.

By setting $c_{1}$ value in (\ref{BTZ2}), $c_{1} = GM_{\text{BTZ}}/4\pi^2$, one can find the holographic entanglement entropy for the solution (\ref{BTZ2}) easily. Let us remind that the holographic entanglement entropy of 2D CFT at finite temperature which is dual to BTZ black hole is
\begin{align}
\label{SEECFT}
S_{EE}^{\text{CFT}}=\frac{c}{3}\log \left(\frac{1}{\epsilon}\frac{l}{\sqrt{2 G M_{\text{BTZ}}}  }\sinh(\frac{R\sqrt{2G
M_{\text{BTZ}}}}{l})\right),
\end{align}
where $R$ is the size of entangling region. Substituting $M_{\text{BTZ}}$ with $M_{T\overline{T}}$ from (\ref{MBTZ}) in (\ref{SEECFT}), the result to first order in $\mu$ becomes
\begin{equation}
\label{SEETTb}
S_{EE}^{T\overline{T}} = S_{EE}^{\text{CFT}}+\mu\frac{\pi c^2} {144 \beta ^2}\left(\frac{\pi R}{\beta
}\coth(\frac{\pi R}{\beta })-1\right),
\end{equation}
where we have also used the Eq.(\ref{MBTZ}). The result  (\ref{SEETTb}) has extra term (last term in big parentheses) in comparison with the previous study \cite{Chen} in which the authors assumed the finite cutoff interpretation. This extra term causes that entanglement entropy (\ref{SEETTb}) vanishes for zero entangling region. To close this section, we report the final result for entanglement entropy of ground state (\ref{FGPS2}) (with $a=1$) \footnote{An easy way to see this result is substituting $M_{\text{BTZ}}$ with $-1/8G$ in (\ref{SEETTb}) together with using (\ref{MBTZ}).}
\begin{equation} 
S_{EE}^{T\overline{T}}\simeq \frac{c}{3} 
\log \left(\frac{R}{\epsilon }\right)+\mu\frac{c^2 R^2}{6912 \pi  l^4},\hspace{.5mm}
\end{equation}
which again it vanishes in the UV limit, $R\rightarrow 0$, same as the result in \cite{Donnelly}. It is worth noting that from this latter result we can not conclude that the theory flows to a trivial theory since we have assumed $\mu$ is small.
\section{Path Integral Complexity}
It seems that the similarity between tensor network representation and Path integral representation of vacuum state, can be utilize to define computational complexity of ground state (\ref{PIn}) as the minimum value of the $S_{GL}[\Omega,g]$ action. But   $S_{GL}[\Omega,g]$ (\ref{SGL}), not only depends to the final metric $e^{2\Omega}g$ but also it depends to the reference metric $g$ which means that this action does not provide us with an absolute quantity which measures the complexity of the optimized state. According to \cite{PI2}, it is convenient to
look at the relative quantity $I_{GL}[{g_2,g_1}]$ which satisfy the following identity
\begin{equation}
\label{Ig1g2}
I_{GL}[g_{1},g_{2}] +I_{GL}[g_2,g_3] = I_{L}[g_1,g_3].
\end{equation}
The above relation implies that
\begin{equation}
I_{GL}[e^{2\Omega}g,\tilde{g}] = I_{GL}[e^{2\Omega}g,g] -I_{GL} [\tilde{g},g],
\end{equation}
which means that $I_{GL}[g_2,g_1]$ actually measures the difference of complexity between the path-integral in $g_2$ and $g_1$. In order to find this quantity, motivating by the form of $S_{GL}$ (\ref{SGL}), we assume the general following form
\begin{align}
\label{Igen}
I_{GL}&[e^{2\Omega}g,g]=\int d^{2}\sigma\sqrt{g}\hspace{.5mm}\bigg(R[g]\Omega+\nabla_{\alpha}\Omega\nabla^{\alpha}\Omega+b_{1}(\Omega)\bigg)+
\nonumber\\
&+\tilde{\mu}\sqrt{g}\hspace{.5mm}\bigg(b_{2}(\Omega)e^{2\Omega}
+b_{3}(\Omega)\Box^{2}\Omega \hspace{1mm}+e^{-2\Omega}b_{4}(\Omega) R^{2}[g]\hspace{.5mm}+
\nonumber\\
&+e^{-2\Omega}R[g]\hspace{.5mm}\big(b_{5}(\Omega)\hspace{.5mm}\Box\Omega+b_{6}(\Omega)\nabla_{\alpha}\Omega\nabla^{\alpha}\Omega\big)\hspace{.5mm}+\nonumber\\
&+(\nabla_{\alpha}\nabla^{\beta}\Omega)\big(b_{7}(\Omega)\nabla^{\alpha}\Omega\nabla_{\beta}\Omega+b_8(\Omega)\nabla_{\beta}\nabla^{\alpha}\Omega\big)\hspace{.5mm}+
\nonumber\\
&+(\nabla_{\beta}\Omega\nabla^{\beta}\Omega)\hspace{.5mm}\big(b_{9}(\Omega)\nabla_{\alpha}\Omega\nabla^{\alpha}\Omega
+b_{10}(\Omega)\hspace{.5mm}\Box\Omega\big)
\bigg).
\end{align}
where $b_{i}$'s are arbitrary function of $\Omega$. Now, the constraint equation (\ref{Ig1g2}) implies that, the  general function (\ref{Igen}) reduces to
\begin{align}
\label{IGL}
&I_{GL}[\Omega,g]=\hspace{-1mm}\int d^{2}\sigma \sqrt{g}\bigg(R[g]\Omega+\nabla_{\alpha}\Omega\nabla^{\alpha}\Omega+b_{1}(e^{2\Omega}-1)\hspace{.5mm}+
\nonumber\\
&\hspace{.7cm}+\tilde{\mu}e^{-2\Omega}\bigg[b_{2}e^{2\Omega}\hspace{.3mm}\big(e^{2\Omega}-1\big)-b_3\hspace{.3mm} \Box\Omega\hspace{.5mm}\big(R[g]-\Box\Omega\big)
\hspace{.5mm}-
\nonumber\\
&\hspace{.7cm}-\dfrac{b_{3}}{4}R^{2}[g]\hspace{.5mm}\big(e^{2\Omega}-1\big)\bigg]\bigg).
\end{align}
Comparing this action with (\ref{SGL}) results in $b_1 =-\tilde{\Lambda}$, $b_2 = -3\tilde{\Lambda}^{2}/16$, $b_3 = 3/16$. 
To have a well defined variational principle, we should add proper Generalized Gibbons-Hawking term to the above action which it is given by
\begin{align}
&I_{GGH}[\Omega,\gamma]=2\hspace{-.5mm}\int \hspace{-.5mm}dx \sqrt{\gamma}\hspace{.5mm}\bigg(K[\gamma]\Omega -\frac{3}{16}\tilde{\mu}e^{-2\Omega} \bigg[ K[\gamma]\hspace{.5mm}\Box\Omega\hspace{.5mm}+
\nonumber\\
&\hspace{.5cm}+\frac{1}{2}K[\gamma]R[g]\big(e^{2\Omega}-1\big)\hspace{.5mm} - n^{\alpha}\nabla_{\alpha}\Omega\hspace{.5mm}\big(R[g]-\Box\Omega\big)\hspace{.5mm} \bigg]\bigg),
\end{align}
where $\gamma$ and $K[\gamma]$ are respectively induce metric and its extrinsic curvature. In the above we have set $\tilde{\Lambda}=1/l^{2}$. Now we have all ingredients to calculate the path integral complexity
\begin{equation}
\mathcal{C}_{T\overline{T}} \equiv \frac{c}{24\pi}\big(I_{GL}[\Omega,g] + I_{GGH}[\Omega,\gamma]\hspace{.5mm}\big), 
\end{equation}
for the solutions which we have found in previous section. The final results for solutions (\ref{PS}) (with $c_1=-1/32\pi^2$) and (\ref{BTZ1}) (with $c_{1} = G M_{\text{BTZ}}/4\pi^{2}$) are respectively \footnote{We would like to mention here that for calculating  $\mathcal{C}_{T\overline{T}}^{\text{BTZ}}$ we have used $1/\beta^{\prime}$, (\ref{TTTb}), not $1/\beta$. If one uses the undeformed temperature, (\ref{MBTZ}), the complexity contains new UV divergence at order $\mathcal{O}(1/\epsilon^2)$ which is clearly in contradiction with holographic complexity of $T\overline{T}$ deformed CFTs.},
\begin{align}
&\mathcal{C}_{T\overline{T}}^{\text{AdS}} = \frac{c}{6}\left(1+\frac{\pi c \mu}{64l^{2}}\right)\frac{l}{\epsilon} -\frac{c}{6}\left(1+ \frac{\pi c \mu}{32l^2}(\frac{1}{3\pi^2}+\frac{1}{2})\right)-
\nonumber\\ 
&\hspace{.22cm}-\frac{\pi c^2 \mu }{192l^2} (\frac{1}{3\pi^2 }+\frac{
1}{2})\hspace{.5mm}\frac{z_{\infty}}{l},
\nonumber\\\nonumber\\
&\mathcal{C}_{T\overline{T}}^{\text{PS}}=\frac{c}{6}\left(1+\frac{\pi c \mu}{64l^{2}}\right)\frac{l}{\epsilon} -\frac{a c}{6}\left(1+ \frac{\pi c \mu}{32l^2}(\frac{1}{3\pi^2}+\frac{1}{2})\right)+
\nonumber\\ 
&\hspace{.22cm}+\frac{c}{12}\left((a^{2}-1)-\frac{\pi c \mu }{16l^2} (\frac{a^{2}}{3\pi^2 }+\frac{
1}{2}) \right)\frac{z_{\infty}}{l},
\nonumber\\\nonumber\\
&\mathcal{C}_{T\overline{T}}^{\text{BTZ}} =\frac{c}{3}(1+\frac{\pi c \mu}{64 l^2})\hspace{.5mm}\frac{l}{\epsilon}\hspace{.5mm}- 
\nonumber\\
&\hspace{.22cm}-\frac{\beta  c}{24 l}-\frac{\pi ^2 c \hspace{.3mm}l}{6\beta}-\frac{\pi c \mu}{6}  \left(\frac{\beta 
c}{128 l^3}+\frac{\pi ^2 c \hspace{.3mm}l}{48\beta^3 
}-\frac{c}{192 \beta l}\right),
\end{align}
where $z_{\infty}(\rightarrow \infty)$ is the IR cut off in the $z$ integral. This IR divergence might be related to the non-local nature of $T\overline{T}$ deformed CFTs \cite{Cardy}. Remarkably, the complexity of formation,   $\mathcal{C}^{\text{BTZ}}-2\hspace{.5mm}\mathcal{C}^{\text{AdS}}$, in these theories is UV finite same as with 2D CFTs. This result that the UV divergence of $\mathcal{C}_{T\overline{T}}^{\text{BTZ}}$ is independent from its mass is in agreement with $\text{CA}$ holographic complexity proposal \footnote{The holographic complexity of formation for BTZ black hole (\ref{BTZ}) can be found in \cite{MyersCF}.}. 


We would like close this section by pointing the interesting consistency of the action (\ref{IGL}). We show that the stress tensor obtained from the first law of entanglement entropy is same as the one obtained from the action (\ref{IGL}). Let us remind that 
the change in entanglement entropy under a
small variation of a quantum state is captured by the variation in the Weyl factor field, $\Omega(z) \rightarrow \Omega_{0}(z)+\delta\Omega(z)$. This fact implies that for a small entangling region $A =[-R/2,R/2]$, the change in entanglement entropy becomes
\begin{align}
\label{deltaSEE}
\Delta S_{A} \simeq \frac{c}{6} \int ds\hspace{.5mm}e^{\Omega(z)} \delta \Omega(z) = \frac{c R^{2}}{24} \partial^{2}_{z}\delta\Omega(z),
\end{align}
where for the thermofield double solution (\ref{BTZ1}) we have
\begin{align}
\label{deltaOmega}
\delta\Omega(z) = \frac{2\pi^{2}}{3\beta^{\prime \hspace{.5mm}2}}z^{2} +\mathcal{O}(z^{4}),
\end{align}
with $\beta^{\prime}$ is given by (\ref{TTTb}). Therefore, according to the first law of entanglement entropy \cite{Takayanagi2012}
\begin{align}
\label{FirstLaw}
\Delta S_{A} \simeq \frac{\pi R^{2}}{3} T_{tt},
\end{align}
the energy density becomes
\begin{align}
\label{Ttt}
T_{tt} = \frac{\pi c}{6\beta^{\prime\hspace{.5mm}2}}.
\end{align}
Independently, by taking the metric variation from the action (\ref{IGL}) and noting that $b_1 =-\tilde{\Lambda}$, $b_2 = -3\tilde{\Lambda}^{2}/16$, $b_3 = 3/16$, the corresponding stress tensor components are given by 
\begin{align}
\label{Tcomp}
&\frac{12\pi}{c}T_{ww}=\partial
^2\Omega-(\partial
\Omega)^2-\frac{3}{4}\tilde{\mu}
e^{-2 \Omega}
\bigg(\partial^3\bar{\partial}\Omega-6\partial \Omega
\partial^2\bar{\partial}\Omega\hspace{.5mm}+
\nonumber\\
&\hspace{1cm}+2\big[4
\left(\partial
\Omega\right)^2-\partial
^2\Omega\big]\partial\bar{\partial}\Omega\bigg),
\nonumber\\
&
\frac{12\pi}{c}\overline{T}_{\bar{w}\bar{w}}=\bar{\partial}
^2\Omega-(\bar{\partial}
\Omega)^2-\frac{3}{4}\tilde{\mu} 
e^{-2 \Omega}
\bigg(\partial\bar{\partial}^3\Omega-6\bar{\partial} \Omega
\partial\bar{\partial}^2\Omega\hspace{.5mm}+
\nonumber\\
&\hspace{1cm}+2\big[4
\left(\bar{\partial}
\Omega\right)^2-\bar{\partial}
^2\Omega\big]\partial\bar{\partial}\Omega\bigg),
\nonumber\\
&\frac{12\pi}{c}T_{w\bar{w}}=\frac{e^{2\Omega}-1}{4
	l^2}-\partial\bar{\partial}\Omega+\frac{3}{64} \tilde{\mu}\hspace{.5mm}
\bigg(\frac{e^{2\Omega}-1}{l^4}
+16e^{-2 \Omega}
\big[
\nonumber\\
&\hspace{1.0cm}\partial^2\bar{\partial}^2\Omega
-2
\partial\Omega\partial\bar{\partial}^2
\Omega-2\bar{\partial}\Omega
\partial^2\bar{\partial}\Omega
-3\left(\partial\bar{\partial}\Omega\right)^2\hspace{.5mm}+
\nonumber\\
&\hspace{1cm}+4\partial\Omega\bar{\partial}\Omega
\partial\bar{\partial}\Omega\big]\bigg),
\end{align}
which for the solution (\ref{BTZ1}), they give together exactly the energy density (\ref{Ttt}). One can also see that for deformed primary states (\ref{PS}), the energy density obtained from the action (\ref{IGL}) becomes
\begin{align}
\label{TTTba}
T_{tt} = -\frac{a^2\hspace{.5mm}c}{24\pi l^2}(1-\frac{\tilde{\mu}}{8\pi^2 l^2}),
\end{align}
which matches with the one obtained from the first law of entanglement entropy (\ref{FirstLaw}). Last but not least, the energy densities (\ref{Ttt}) and (\ref{TTTba}) are in complete agreement with the variational principle analysis of \cite{Guica}. The authors of \cite{Guica} have shown that if a 2D CFT lives on the conformal boundary $g_{\alpha\beta}$ in (\ref{FG}), the deformed stress tensor becomes
\begin{align}
\label{TGuica}
T^{[\mu]}_{\alpha\beta} = \frac{1}{8\pi G l}\hspace{.5mm} g^{(2)}_{\alpha\beta} +\frac{\mu}{64 \pi^2 G^2 l^2}\hspace{.5mm} g^{(4)}_{\alpha\beta}.
\end{align}
For example for the BTZ solution, $g_{tt}^{(2)} = 2\pi^2 l^2/\beta^2$, $g_{tt}^{(4)} = \pi^4 l^4/\beta^4$ and  $c=3l/2G$ which they together imply that the deformed energy density (\ref{TGuica}) exactly matches with (\ref{Ttt}).
More intriguingly, the deformed energy density (\ref{Ttt}), for positive value of $\mu$, is nothing just the $tt$-component of Brown-York stress tensor on the $\rho=\rho_{c}$ surface
with $\rho_{c}$ is given by (\ref{rcut}). But without any need to this bulk cut off interpretation, the stress tensor (\ref{Tcomp}) is conserved and it satisfies the Zamolodchikov's flow equation.
\vskip.3cm
\section{Summary and Outlook}
We have shown that the path integral optimization approach, for $T\overline{T}$ deformation of 2D CFTs,  implies that the time slice of optimized geometries indeed capture the entire of spacetime. For positive deformation coupling, these optimized solutions can be reinterpreted as a geometry at finite cut-off radius. In this letter we studied the entanglement entropy and quantum complexity for those optimized solutions. Another interesting quantities which should be studied are correlation functions. Investigating the relation between entanglement of purification and holographic entanglement wedge cross section in this context is also an intriguing future problem. 
\vskip.3cm
\section{ACKNOWLEDGMENTS}
We wish to thank Mohsen Alishahiha, Mehregan Doroudiani, Thomas Hartman, Tadashi Takayanagi and Alexander B. Zamolodchikov for fruitful discussions. We are also very grateful to Pawel Caputa for carefully reading the draft and for his valuable comments. This article is part of the PhD project of Hamed Zolfi under the joint supervision of Vahid Karimipour from Sharif University of Technology and Ali Naseh from IPM.

\nopagebreak
\end{document}